\documentclass[aps,reprint,superscriptaddress,nofootinbib,showpacs]{revtex4-1}
\usepackage{amsmath,latexsym,amssymb,hyperref,graphicx}
\usepackage{slashed}

\hypersetup{colorlinks,citecolor= nicegreen,linkcolor= niceblue}
\usepackage{color}
\definecolor{nicered}{rgb}{0.7,0.1,0.1}
\definecolor{nicegreen}{rgb}{0.1,0.5,0.1}
\definecolor{niceblue}{rgb}{0.1,0.1,0.8}

\newcommand{\beq}{\begin{equation}}
\newcommand{\eeq}{\end{equation}}
\newcommand{\bea}{\begin{eqnarray}}
\newcommand{\eea}{\end{eqnarray}}

\begin{document}

\title{Gravitational localization of scalar zero modes in $SU(5)\times Z_2$ branes}

\author{Nelson Pantoja}
\affiliation{Centro de F\'isica Fundamental, Universidad de Los Andes,
M\'erida, Venezuela}

\author{Rossana Rojas}
\affiliation{Centro de F\'isica Fundamental, Universidad de Los Andes,
M\'erida, Venezuela}

\date{\today}

\begin{abstract}
  \noindent  
The fluctuations of 5D self-gravitating non-abelian kinks which arise from the breaking of the $SU(5)\times Z_2$ symmetric theory are analyzed within the context of brane-worlds. While tensor and vector sectors of these fluctuations behave like its counterparts in the standard abelian $Z_2$ kinks, the mixing between the field excitations of the non-abelian kink and the scalar components of the metric makes the pure scalar sector of the theory very interesting. The spectrum of these scalar fluctuations, which includes gravitationally trapped massless modes on the core of the wall associated to the broken symmetries, is discussed for the two classes of kinks that break $SU(5)$ to its maximal subgroups.
\end{abstract}


\maketitle

\section{Introduction}

The central idea in brane-world scenarios is that matter and its interactions, as we know them, must be localized on the brane. Domain wall spacetimes, providing regularizations of the Randall-Sundrum brane-world \cite{Randall:1999vf,Guerrero:2002ki} preserve the property of produce effective 4D gravity on the brane \cite{DeWolfe:1999cp,Gremm:1999pj,Kakushadze:2000zn,CastilloFelisola:2004eg}. In addition, while it has been shown that fermion fields may be localized on the brane {\it via} their interaction with the scalar field from which the wall is made of \cite{Bajc:1999mh,Kehagias:2000au,Ringeval:2001cq,Koley:2004at,Melfo:2006hh,Liu:2009dw,Guerrero:2006gj}, gauge field localization has been somewhat more elusive \cite{Huber:2000fh,Dvali:2000rx,Dubovsky:2001pe,Batell:2006dp,Guerrero:2009ac,Chumbes:2011zt,Germani:2011cv}. 

A lot of work has been done on the topic of localization in brane-worlds 
in which the field theoretic domain walls considered arise in abelian 
$Z_2$-symmetric theories, although it is 
also possible to consider (besides the $Z_2$ symmetry) continuos 
internal symmetry walls with interesting localization properties. 
There are well known examples of domain walls in flat space-time  
that arise in a  $SU(5)\times Z_2$ symmetric theory 
\cite{Dvali:1997sa,Pogosian:2000xv,Vachaspati:2001pw,Pogosian:2001fm} 
in which the full $SU(5)\times Z_2$ 
symmetry is not restored at the core of the wall \cite{Vachaspati:2003zp}. 
This makes 
non-abelian domain walls of this sort (rather its extensions to the 
gravitating case) very interesting within the context of brane-worlds 
\cite{Vachaspati:2003zp,Davidson:2002eu,Davidson:2007cf,Melfo:2011ev}. 

It has been shown that non-abelian branes in flat space that break 
$E_6$ to $SO(10)\times U(1)$ may produce, {\it via} the 
Dvali-Shifman (DS) mechanism \cite{Dvali:1996xe}, an $SU(5)$ effective 
theory on the brane \cite{Davidson:2007cf}.
On the other hand, self-gravitating $SU(5)\times Z_2$ domain walls, 
supporting 4D massless fermion excitations in its world volume, 
may also localize 4D massless non-abelian excitations via the DS mechanism, 
depending of the symmetry breaking pattern \cite{Melfo:2011ev}.

While domain walls in abelian $Z_2$-symmetric theories are 
topologically stable, there is no global stability criterium for the non-abelian 
ones. This lead us to resort to perturbative analysis, after a solution is 
found, to establish at least their local stability \cite{Vachaspati:2001pw,Pogosian:2001fm}. Furthermore, the inclusion of gravity makes the 
perturbative analysis somewhat more intricate than the analogous one 
in flat spacetime due to the mixing between the field excitations of the 
non-abelian kink $\Phi$ and the scalar components of the metric $g_{ab}$, 
in a theory invariant under diffeomorphisms acting both on the 
$\Phi$ and on $g_{ab}$. 

In this work we make a perturbative analysis for the 5D self-gravitating 
$SU(5)\times Z_2$ domain walls of \cite{Melfo:2011ev} in terms of 
diffeomorphism-invariant quantities. This analysis not only shows that 
these walls are perturbatively stable, it also permits to study the 
gravitationally trapped content from the point of view of 4D observers 
localized on the brane.  We find, as expected, that the 
tensor and vector sectors of the fluctuations behave as its counterparts in 
the more familiar $Z_2$-symmetric domain walls. On the 
other hand, the scalar sector of the fluctuations in the $SU(5)\times Z_2$ 
case shows a rather different behavior from the abelian one, being linked 
to the particular symmetry breaking pattern considered and 
with a spectrum that includes normalizable massless modes, i.e., with 
massless scalar particles associated to the broken symmetries which are gravitationally trapped on the core of the wall.

\section{Self-gravitating $SU(5)\times Z_2$ kinks}\label{SU_5_branes}

In this section we briefly recall the results of \cite{Melfo:2011ev} regarding 
the properties of 5D selfgravitating domain walls formed in the two possible 
symmetry breaking patterns of $SU(5)\times Z_2$. Let us consider the 
$(4+1)$-dimensional theory 
\begin{equation}\label{theory}
S=\int d^4x\,d\xi\, \sqrt{-g}\left[\frac{1}{2}R-\text{Tr}(\partial_a\Phi \partial^a\Phi) - V(\Phi)\right],
\end{equation}
where $R$ is the scalar curvature, $g$ is the determinant of the metric, $\Phi$ is a scalar field that transforms in the adjoint representation of $SU(5)$ and $V(\Phi)$ a potential such that (\ref{theory}) is invariant under the transformations 
                         \begin{itemize}
                         \item {$\Phi\longrightarrow U\Phi U^\dagger$},  {$U=\exp\{i\omega_q\mathbf{T}^q\}$},
                         \item {$Z_2: \Phi \longrightarrow -\Phi$}, {$Z_2 \notin SU(5)$} 
                         \end{itemize}
where $\mathbf{T}^q, \,\,q=1,\ldots ,24$, are traceless hermitian generators of $SU(5)$.\footnote{We use units were $\hbar = G = c = 1$} 

The Einstein-scalar field equations for this system are
\begin{equation}\label{field_eq1}
 R_{ab}-\frac{1}{2}g_{ab}R=T_{ab},
\end{equation}
where
\begin{equation}\label{field_eq2}
T_{ab}=2\text{Tr}\left(\nabla_{a}\Phi\nabla_{b}\Phi\right)-g_{ab}\left(g^{cd}\text{Tr}\left(\nabla_{c}\Phi\nabla_{d}\Phi\right)+V(\Phi)\right)
\end{equation}
and
\begin{equation}\label{field_eq3}
g^{ab}\nabla_{a}\nabla_{b}\phi_m=\frac{\partial V(\Phi)}{\partial\phi_m},\quad \Phi=\phi_m\mathbf{T}^m,
\end{equation}
where $\nabla_{c}g_{ab}=0$. Exact domain wall solutions of (\ref{field_eq1},\ref{field_eq2},\ref{field_eq3}) are available for a sixth-order Higgs potential of the form
\begin{eqnarray}\label{potential1}
V(\Phi) \!\!&=&\!\!V_0-\mu^2 \text{Tr}[\Phi^2]+h (\text{Tr}[\Phi^2])^2+\lambda \text{Tr}[\Phi^4] \nonumber\\&&\!\!+ \alpha (\text{Tr}[\Phi^2])^3 + \beta (\text{Tr} [\Phi^3])^2+ \gamma (\text{Tr} \Phi^4)(\text{Tr} \Phi^2),
\end{eqnarray}
for special values of the couplings which yield integrable models.

Assuming that the geometry preserves 4D-Poincare invariance, the metric {\it ansatz} is
\begin{equation}
g_{ab}= e^{2A(\xi)}\eta_{\mu\nu}dx^{\mu}_adx^{\nu}_b + d\xi_a d\xi_b,
\end{equation}
with $\eta_{\mu\nu}=\text{diag}(-1,+1,+1,+1)$ and a domain wall 
solution $\Phi$ of the form
\begin{equation}
\Phi(\xi)=\phi_M(\xi) \mathbf{M} + \phi_P (\xi) \mathbf{P},
\end{equation}
is proposed, where $\phi_M,\phi_P$ satisfy the boundary conditions
\begin{equation}
\phi_M(+\infty)=-\phi_M(-\infty),\qquad \phi_P(+\infty)=\phi_P(-\infty),
\end{equation}
with $\mathbf{M}$ and $\mathbf{P}$ appropriately chosen orthogonal diagonal generators of $SU(5)$. 

Imposing the integrability conditions 
\begin{equation}\label{integrability1}
\text{Tr}(\mathbf{M}\mathbf{P})=\text{Tr}(\mathbf{M}^2 \mathbf{P})= \text{Tr}( \mathbf{M}^3 \mathbf{P}) = 0,
\end{equation}
\begin{equation}\label{integrability2}
\text{Tr}(\mathbf{P}^3 \mathbf{M})=
\text{Tr}(\mathbf{P}^3) =0, 
\end{equation}
for the special values of the couplings given by 
\begin{eqnarray} \label{conditions}
 h &=& - 12 \text{Tr}(\mathbf{P}^2 \mathbf{M}^2) \,  \lambda, \nonumber \\
 \alpha &=& \frac{4}{3}\left[ \frac{ 2 \text{Tr} (\mathbf{M}^3) \text{Tr} (\mathbf{P}^4) -    3 \text{Tr} (\mathbf{P}^2 \mathbf{M})  \text{Tr} (\mathbf{M}^4) }{ 3 \text{Tr} (\mathbf{P}^2 \mathbf{M})  - 2 \text{Tr} (\mathbf{M}^3) }\right.\nonumber\\
 &&\left. \qquad-  6 \text{Tr}(\mathbf{P}^2 \mathbf{M}^2) \right] \, \gamma, \nonumber \\
  \beta &=&  \frac{1}{6}\left[  \frac{\text{Tr} (\mathbf{M}^4) - \text{Tr} (\mathbf{P}^4)}{ \text{Tr} (\mathbf{P}^2 \mathbf{M}) [3 \text{Tr} (\mathbf{P}^2 \mathbf{M})  - 2 \text{Tr} (\mathbf{M}^3) ]}   \right] \, \gamma,
\end{eqnarray}
an exact domain wall solution is given by \cite{Melfo:2011ev} 
\begin{equation}
 \phi_M(\xi)= v\tanh b\xi,\quad \phi_P(\xi)= v\kappa
\end{equation}
where $\mu^2$, $\lambda$ y $\gamma$ can be written explicitly in terms of $v$ and $b$, and $\kappa$ is a numerical constant which depends on the choice of $\mathbf{M}$ and $\mathbf{P}$. 
On the other hand,
\begin{equation}
 A(\xi)=-\frac{v^2}{9}[2 \ln \left(\text{cosh}(b\xi) \right)+\frac{1}{2}\tanh^2(b\xi)]
 \label{warp}
\end{equation}
and the space-time is asymptotically $AdS_5$ with cosmological constant $\Lambda=V(\Phi)_{\xi=-\infty}=V(\Phi)_{\xi=+\infty}=-\frac{8}{27}b^2 v^4$.

As discussed in \cite{Melfo:2011ev} (see \cite{Vachaspati:2001pw,Pogosian:2001fm} for the flat space case), 
the choice of $\mathbf{M}$ and $\mathbf{P}$ relies on the desired 
asymptotic values for $\Phi$ at $\xi\rightarrow \pm\infty$.  
It is well known that in flat space-time there 
are two symmetry breaking patterns of $SU(5)$ by a field $\Phi$ in the 
adjoint, involving  the breaking to subgroups with the same rank as 
$SU(5)$ and occurring as minima of a fourth-order potential 
\cite{Li:1973mq,Chen:2010er}. These are 
\begin{equation}\label{first}
SU(5) \times Z_2 \rightarrow {SU(3)\times SU(2) \times U(1)}/({Z_3\times Z_2})\end{equation}
and 
\begin{equation}\label{second}
SU(5) \times Z_2 \rightarrow {SU(4) \times U(1)}/{Z_4},
\end{equation}
the second type of symmetry breaking pattern yielding the largest residual 
symmetry. In the remainder of this section, we quote the results of \cite{Melfo:2011ev} concerning to the 
symmetry breakings (\ref{first}) and (\ref{second}) for the gravitating case 
with the sixth order potential and refer the reader to that work for a detailed discussion on these and the 
other issues related to localization of fermions and gauge fields within the 
context of brane-worlds. 

\subsection{Symmetry breaking $SU(5) \times Z_2 \rightarrow {SU(3)\times SU(2) \times U(1)}/({Z_3\times Z_2})$}
For this symmetry breaking pattern we have
  \begin{equation}
\mathbf{M}_{\bf A} =\frac{1}{\sqrt{40}} {\rm diag (1,1,1,1,-4)},
\end{equation}
\begin{equation}
\mathbf{P}_{\bf A} =\frac{1}{2\sqrt{2}}  {\rm diag (1,1,-1,-1,0)}
\end{equation}
and $\kappa_A=\sqrt{5}$. With $H_{\pm}$ and $H_0=H_+\cap H_-$ the unbroken symmetries at $\xi\rightarrow \pm\infty$ and $\xi=0$, respectively, we have 
\begin{equation}
H^{\bf A}_\pm = \frac{SU(3)_\pm \times SU(2)_\pm \times U(1)_\pm}{Z_2\times Z_3},
\end{equation}
\begin{equation}
 H^{\bf A}_0 = \frac{SU(2)_+ \times SU(2)_- \times U(1)_M \times U(1)_P}{Z_2 \times Z_2},
\end{equation}
with the following embeddings
\begin{equation}
SU(2)_\pm \subset SU(3)_\mp,
\end{equation} 
in the sense that the Cartan subalgebra de $SU(2)_+$ ($SU(2)_-$) is a 
subspace of the  Cartan-subalgebra space of $SU(3)_-$ ($SU(3)_+$) 
corresponding to the particular basis chosen for the Lie algebra. 
$H^A_0$ resembles the gauge group 
$SU(2)_L\times SU(2)_R\times U(1)_{B-L}$ on which is based the 
Higgs sector of the minimal 4D Left-Right symmetric theory. This 5D 
non-abelian wall may be thought as the generalization to the gravitating 
case of the perturbatively stable flat space one for a quartic potential of 
the references \cite{Pogosian:2000xv,Vachaspati:2001pw,Pogosian:2001fm}.

\subsection { Symmetry breaking $SU(5) \times Z_2 \rightarrow {SU(4) \times U(1)}/{Z_4}$}

In this case,
 \begin{equation}
\mathbf{M}_{\bf B}=\frac{1}{\sqrt{60}} {\rm diag (-2,-2,-2,3,3)}
\end{equation}
\begin{equation}
\mathbf{P}_{\bf B}=\frac{1}{2}  {\rm diag (0,0,0,1,-1)}
\end{equation}
and $\kappa_B=\sqrt{5/3}$. The unbroken symmetries are
\begin{equation}
 H^{\bf B}_\pm = \frac{SU(4)_\pm \times  U(1)_\pm}{Z_4}
\end{equation}
 \begin{equation}
 H^{\bf B}_0 = \frac{SU(3) \times U(1)_M \times U(1)_P}{Z_3 }
 \end{equation}
 where {$SU(3)$} is embedded in diferent manners in {$SU(4)_+$} and 
 {$SU(4)_-$}. In this symmetry breaking pattern, the $SU(3)$ gauge bosons 
 on the wall need to form massive glueballs of $SU(4)$ in order to escape 
 into the bulk, so they can be localized on the wall {\it via} the 
 Dvali-Shifman mechanism \cite{Dvali:1996xe}.

\section{Diffeomorphism-invariant fluctuations}

Let $g_{ab}$ and $\Phi=\phi_q\mathbf{T}^q$ be the metric and the scalar field solutions of the field equations of the theory (\ref{theory}). Let $h_{ab}$ and $\boldsymbol{\varphi}=\varphi_q \mathbf{T}^q$ the metric and scalar field fluctuations, respectively, around the above background.
 Following \cite{CastilloFelisola:2004eg} we find that these fluctuations satisfy
\begin{eqnarray}\label{fluctuations1}
-\frac{1}{2}g^{cd}\nabla_{c}\nabla_{d}h_{ab}+R^{c}{}_{(ab)}{}^{d}h_{cd}+R^{c}_{(a}h^{ }_{b)c}\nonumber\\
-\frac{1}{2}\nabla_{a}\nabla_{b}\left(g^{cd}h_{cd}\right)+\nabla_{(a}\nabla^{c}h_{b)c}=\nonumber\\\!\!\!\!4\text{Tr}[\nabla_{(a}\Phi\nabla_{b)}\boldsymbol{\varphi}] -\frac{2}{3}h_{ab}V(\Phi)-\frac{2}{3}\!\!\left(\frac{\partial V(\Phi)}{\partial \phi_q}\varphi_q \right) \!\!g_{ab}
\end{eqnarray}
and
\begin{eqnarray}\label{fluctuations2}
\!\!\!\!-h^{ab}\nabla_{a}\nabla_{b}&\phi_q&-\frac{1}{2}g^{cd}\left(\nabla_{a}h_{bd}+\nabla_{b}h_{ad}-\nabla_{d}h_{ab}\right)\nabla_{c}\phi_q \nonumber\\
\!\!\!\!&+&g^{ab}\nabla_{a}\nabla_{b}\varphi_q-\frac{\partial^2 V(\Phi)}{\partial\phi_p\partial\phi_q}\varphi_p=0,
\end{eqnarray}
where $\nabla_{c}g_{ab}=0$ and
\begin{equation}
V(\Phi +\boldsymbol{\varphi})=V(\Phi) + \frac{\partial V(\Phi)}{\partial\phi_q}\varphi_q + \frac{1}{2}\frac{\partial^2 V(\Phi)}{\partial\phi_p\partial\phi_q}\varphi_p\varphi_q + O(\boldsymbol{\varphi}^3).
\end{equation}

For the theory (\ref{theory}) with the sixth order potential 
(\ref{potential1}) we have
\begin{eqnarray}\label{delta_V}
\frac{\partial V(\Phi)}{\partial\phi_q}&=&\left(-2\mu^2 + 4h\text{Tr}[\Phi^2] + 6 \alpha(\text{Tr}[\Phi^2])^2 \right.\nonumber\\&&\left.+ 2\gamma\text{Tr}[\Phi^4]\right)\text{Tr}[\Phi \mathbf{T}^q]+ 6\beta\text{Tr}[\Phi^3]\text{Tr}[\Phi^2 \mathbf{T}^q] \nonumber\\&&+ 4\left(\lambda +\gamma\text{Tr}[\Phi^2]\right)\text{Tr}[\Phi^3 \mathbf{T}^q]
\end{eqnarray}
and
\begin{eqnarray}\label{delta_V_2}
\frac{\partial^2 V}{\partial\phi_q\partial\phi_p}=&&\left(-2\mu^2 + 4h\text{Tr}[\Phi^2] + 6\alpha(\text{Tr}[\Phi^2])^2 \right.\nonumber\\&&\left.+2\gamma\text{Tr}[\phi^4]\right)\text{Tr}[\mathbf{T}^q\mathbf{T}^p]+ 6\beta\left(\text{Tr}[\Phi^3](\text{Tr}[\mathbf{T}^q\mathbf{T}^p\Phi] \right.\nonumber\\
&&\left.+ \text{Tr}[\mathbf{T}^p\mathbf{T}^q\Phi]) + 3\text{Tr}[\mathbf{T}^q\Phi^2]\text{Tr}[\mathbf{T}^p\Phi^2]\right)\nonumber\\
&& + 4(\lambda +\gamma\text{Tr}[\Phi^2])\left(\text{Tr}[\mathbf{T}^q\mathbf{T}^p\Phi^2] + \text{Tr}[\mathbf{T}^p\mathbf{T}^q\Phi^2] \right.\nonumber\\
&&\left.+\text{Tr}[\mathbf{T}^q\Phi\mathbf{T}^p\Phi]\right) + 8(h + 3\alpha\text{Tr}[\Phi^2])\text{Tr}[\mathbf{T}^q\Phi]\text{Tr}[\mathbf{T}^p\Phi]\nonumber\\&& +8\gamma\left( \text{Tr}[\mathbf{T}^q\Phi^3]\text{Tr}[\mathbf{T}^p\Phi] + \text{Tr}[\mathbf{T}^p\Phi^3]\text{Tr}[\mathbf{T}^q\Phi] \right).
\end{eqnarray}
In the background $\{g_{ab}, {\Phi}\}$ provided by the $SU(5)\times Z_2$ 
kinks discussed in the previous section, (\ref{delta_V}) reduces to
\begin{equation}
\frac{\partial V(\Phi)}{\partial\phi_q}= \left(\phi_M'' + 4A'\phi_M'\right)\delta^{Mq}
\end{equation}
and the right hand side of (\ref{delta_V_2}) turns out diagonal in $q,p$. 
However, although they get notably simplified, (\ref{fluctuations1}) 
and (\ref{fluctuations2}) are still extremely involved.

Indeed, as is well known, $h_{ab}$ and $\boldsymbol{\varphi}$ are not diffeomorphism invariant variables. Under an infinitesimal diffeomorphism of the form
\begin{equation}\label{difeomorphism}
 x^{a}\rightarrow\overline{x}^{a}=x^{a}+\epsilon^{a},
 \end{equation}
we have
\begin{equation}\label{change1}
h_{ab}\rightarrow \overline{h}_{ab}=h_{ab}-2\nabla_{(a}\epsilon_{b)}
\end{equation}
and
\begin{equation}\label{change2}
\boldsymbol{\varphi}\rightarrow \overline{\boldsymbol{\varphi}}=\boldsymbol{\varphi}-\epsilon^{a}\nabla_{a}\Phi,
\end{equation}
where $\{h_{ab},\boldsymbol{\varphi}\}$ and $\{\overline{h}_{ab}, \overline{\boldsymbol{\varphi}}\}$ describe the same physical perturbations. 
Hence we must take care of the general coordinate invariance of the theory 
(\ref{theory}).
 
Since the background $\{g_{ab}, {\Phi}\}$ provided by the 
$SU(5)\times Z_2$ kinks preserves 4D-Poincare invariance, we 
decompose $h_{ab}$ (from the point of view of 
the 4-dimensional observers confined to the wall) in scalar, vector 
and tensor sectors. Following \cite{Giovannini:2001fh}, we set
\begin{equation}
h_{\mu\nu}=2e^{2A}\left(h_{\mu\nu}^{TT}+\partial_{(\mu}f_{\nu)}+\eta_{\mu\nu}\psi+\partial_{\mu}\partial_{\nu}E\right),
\end{equation}
\begin{equation}
h_{\mu\xi}=h_{\xi\mu}=e^A\left(D_{\mu}+\partial_{\mu}C\right),
\end{equation}
and
\begin{equation}
h_{\xi\xi}=2\omega,
\end{equation}
where
\begin{equation}
h_{\mu}^{TT}{}^{\mu}=0,\quad
\partial^{\mu}h_{\mu\nu}^{TT}=0
\end{equation}
and
\begin{equation}
\partial^{\mu}f_{\mu}=0,\quad
\partial^{\mu}D_{\mu}=0.
\end{equation}

From (\ref{change1}), with
\begin{equation}
\epsilon_{a}=(e^{2A}\epsilon_{\mu},\epsilon_{\xi})
\end{equation}
where
\begin{equation}
\epsilon_{\mu}=\partial_{\mu}\epsilon+\zeta_{\mu},\qquad
\partial^{\mu}\zeta_{\mu}=0,
\end{equation}
we have that under an infinitesimal diffeomorphism
\begin{equation}
\psi\rightarrow\overline{\psi}=\psi-A'\epsilon_{\xi},\qquad\omega\rightarrow\overline{\omega}=\omega+\epsilon_{\xi}',
\end{equation}
\begin{equation}
E\rightarrow\overline{E}=E-\epsilon,\qquad
C\rightarrow\overline{C}=C-e^{A}\epsilon'+e^{-A}\epsilon_{\xi},
\end{equation}
\begin{equation}
f_\mu\rightarrow\overline{f}_{\mu}=f_{\mu}-\zeta_{\mu},\qquad D_\mu\rightarrow\overline{D}_{\mu}=D_{\mu}-e^{A}\zeta'_{\mu},
\end{equation}
and
\begin{equation}
\overline{h}_{\mu\nu}^{TT}=h_{\mu\nu}^{TT},
\end{equation}
and from (\ref{change2}) we have
\begin{equation}
\overline{\boldsymbol{\varphi}}=\boldsymbol{\varphi}-\Phi'\epsilon_{\xi},\quad \mbox{since}\quad \Phi=\Phi(\xi).
\end{equation}

The $h_{\mu\nu}^{TT}$ sector is automatically diffeomorphism invariant, 
i.e. $\overline{h}_{\mu\nu}^{TT}=h_{\mu\nu}^{TT}$, while it is also 
possible to define an invariant divergenceless vector (since we have 
one vector gauge 
function $\zeta_{\mu}$) 
\begin{equation}\label{invariant_vector}
{V}_{\mu}={D}_{\mu}-e^{A}{f}'_{\mu},
\end{equation}
and two diffeomorphism invariant scalar fluctuations (since we have two gauge functions $\epsilon$ y $\epsilon_{\xi}$) given by
\begin{equation}
{\Gamma}={\psi}-A'\left(e^{2A}{E}'-e^{A}{C}\right)
\end{equation}
and
\begin{equation}
{\Theta}={\omega} + \left(e^{2A}{E}'-e^{A}{C}\right)',
\end{equation}
such that $\overline{{V}}_\mu={V}_\mu$, $\overline{\Gamma}=\Gamma$ and
$\overline{\Theta}=\Theta$.  

On the other hand for the diffeomorphism invariant non-abelian scalar 
field fluctuation we find
\begin{equation}
\boldsymbol{{\chi}}=\boldsymbol{{\varphi}}-\Phi'(e^{2A}{E}'-e^{A}{C})
\end{equation}
such that $\overline{\boldsymbol{\chi}}=\boldsymbol{\chi}$. 

In the (generalized) longitudinal gauge \cite{Giovannini:2001fh}, $E=C=0$ and 
$f_\mu=0$,  the freedom of the coordinate transformations 
(\ref{difeomorphism}) is completely fixed and we have $\Gamma=\psi$, 
$\Theta=\omega$, $V_\mu=D_\mu$ and $\boldsymbol{{\chi}}=
\boldsymbol{{\varphi}}$. This leaves us with the  tensor $\{h_{\mu\nu}^{TT}\}$, 
vector $\{V_\mu\}$ and scalar $\{\Gamma,\Theta, \boldsymbol{{\chi}}\}$ 
sectors which decouple from each other at the linearized level.
 
The tensor and vector sectors of the fluctuations behave as 
their corresponding analogous in the standard abelian $Z_2$ 
kinks, widely discussed in the literature. It is the scalar sector 
of the theory under consideration that is quite 
different from the $Z_2$ case. Nevertheless, in order to render this work 
self-consistent, we will also discuss briefly the tensor and vector 
fluctuations. In the following, the results will be expressed in the conformal 
coordinate
\begin{equation}\label{xi_zeta}
z=\int d\xi\,e^{-A(\xi)}
\end{equation}
such that
\begin{equation}
g_{ab}=e^{2A(z)}\left(\eta_{\mu\nu}dx^{\mu}_adx^{\nu}_b +dz_a dz_b\right).
\end{equation}

\subsection{Tensor fluctuations}
    In the tensor sector $h^{TT}_{\mu\nu}$ with $\Psi_{\mu\nu}\equiv e^{3A/2}h_{\mu\nu}^{TT}$, the modes $\Psi_{\mu\nu}(x,z)\sim e^{ip\cdot x}\Psi_{\mu\nu}(z)$, satisfy the Schr\"odinger-like equation
    \begin{equation}
   \left(-\partial_{z}^{2}+V_{Q_1}\right)\Psi_{\mu\nu}(z)=m^{2}\Psi_{\mu\nu}(z),
   \end{equation}
where 
   \begin{equation}\label{VQ1}
V_{Q_1}=\frac{9}{4}A'^{2}+\frac{3}{2}A'' 
    \end{equation}
and  $p^{\mu}p_{\mu}=-m^{2}$.   $V_{Q_1}$ supports a massless bound 
state which can be identified with the graviton and a tower of 
non-normalizable massive states which propagate in the bulk. As is well 
known, the above equation can be factorized as
\begin{equation}
\left(-\partial_z^2 + V_{Q_1}(z)\right)=\left(\partial_{z} +
\frac{3}{2}A'(z)\right)\left(-\partial_{z} +
\frac{3}{2}A'(z)\right)
\end{equation}
which implies the absence of modes with $m^2<0$, ensuring the stability 
of the system under tensor perturbations. As in the abelian $Z_2$ kink 
\cite{Callin:2004py}, the normalizable massless tensor mode is given by 
$\Psi^0\propto \exp\{3A(z)/2\}$ which reproduces 4D gravity on the core 
of the wall, while the continuum of massive modes gives small corrections 
to this behavior at short distances. 

\subsection{Vector fluctuations}
The vector sector $V_\mu$ satisfies the equations
    \begin{equation}
    \left(\partial_{z}+3A' \right)V_{\mu}(x,z)=0
    \end{equation}
and
    \begin{equation}
    \partial^{\beta}\partial_{\beta}V_{\mu}(x,z)=0,
    \end{equation}
whose solution is given by $V_{\mu}(x,z)=e^{-3A(z)}V_\mu (x)$, with 
$\partial^{\beta}\partial_{\beta}V_{\mu}(x)=0$. On the other hand, from 
(\ref{invariant_vector}) we see that $V_\mu(x,z)$ must be an odd 
function of $z$ and hence $V_\mu(x)=0$. As in the abelian $Z_2$ kink 
\cite{Giovannini:2001fh}, there are no massless vector fluctuations localized 
on the wall.

\subsection{Scalar fluctuations}   
In the scalar sector we have the set of diffeomorphism-invariant fluctuations 
$\{\Gamma, \Theta, \boldsymbol{{\chi}}\}$, which are subject to two constraints. The first one only involves scalars without charge under $SU(5)$ and is given by
\begin{equation}\label{constraint1}
2\Gamma + \Theta=0,
\end{equation}
while the second one involves also the component 
$\chi_M$ of $\boldsymbol{{\chi}}$
\begin{equation}\label{constraint2}
3A'\Theta -3\Gamma' - \phi_M'\chi_M=0.   
 \end{equation}
 From (\ref{constraint1}) and (\ref{constraint2}) it follows that the fluctuations $\Gamma$, $\Theta$ and $\chi_M$ are not independent and correspond to a single physical scalar fluctuation. 
 
With $\Omega\equiv e^{3A/2}\Gamma/\phi_M'$, the modes 
$\Omega(x,z)\sim e^{ip\cdot x}\Omega(z)$ satisfy
    \begin{equation}
    (-\partial_{z}^{2}+V_{Q_{2}})\Omega(z)=m^{2}\Omega(z),
    \end{equation}
    where $V_{Q_{2}}$ is given by
    \begin{equation}
V_{Q_{2}}=-\frac{5}{2}A''+\frac{9}{4}A'^{2}+A'\frac{\phi_M''}{\phi_M'}+2\left(\frac{\phi_M''}{\phi_M'}\right)^{2}-\frac{\phi_M'''}{\phi_M'}.
    \end{equation}
$V_{Q_{2}}$  does not support normalizable massless states. Notice that
    \begin{equation}\label{super1}
    (-\partial_{z}^{2}+V_{Q_{2}})=\left(-\partial_{z}+ \frac{Z'}{Z}\right)\left(\partial_{z}+\frac{Z'}{Z}\right),
    \end{equation}
where
\begin{equation}\label{Z}
Z=e^{3A/2}\frac{\phi'_M}{A'},
\end{equation}
implying the absence of modes with $m^2<0$. This scalar fluctuation has an exact analogous one in the abelian $Z_2$ selfgravitating kink (see \cite{Giovannini:2001fh}). 

Now, let us consider the scalar fluctuation 
$\boldsymbol{{\chi}}=\chi_q\mathbf{T}^q$. Let 
$\boldsymbol{\Xi}=\Xi_q \mathbf{T}^q$ with
\begin{equation}
\Xi_q\equiv e^{3A/2}\left(\chi_q-\frac{\Gamma}{A'}\phi_q'\right).
\end{equation}
The modes $\Xi_q(x,z)\sim e^{ip\cdot x}\Xi_q(z)$ satisfy 
Schr\"odinger-like equations which depends on $q$.
 
First, let us consider perturbations along the $\mathbf{M}$ direction. For $q=M$, (it should be recalled that the fluctuations $\Gamma$, $\Theta$ 
and $\chi_M$ are not independent) $\Xi_M(z)$ satisfy
    \begin{equation}\label{M_mode}
    (-\partial_{z}^{2}+V_{{M}})\Xi_M(z)=m^{2}\Xi_M(z),
    \end{equation}
where $V_{{M}}$ is given by
    \begin{eqnarray}
V_{{M}}=&-&\frac{A'''}{A'}-\frac{3}{2}A''+\frac{9}{4}A'^{2}+ 2\frac{A''^2}{A'^2}
\nonumber\\&+&\left(3A'-2\frac{A''}{A'}\right)\frac{\phi_M''}{\phi_M'}+\frac{\phi_M'''}{\phi_M'}. 
    \end{eqnarray}
In terms of $Z$, as defined in (\ref{Z}), (\ref{M_mode}) can be rewritten as
\begin{equation}\label{super2}
(-\partial_{z}^{2}+V_{M})=\left(\partial_{z}+\frac{Z'}{Z}\right)\left(-\partial_{z} + \frac{Z'}{Z}\right),
\end{equation}
and from (\ref{super1}) and (\ref{super2}), it follows that the eigenvalues 
$m^2$ of the modes of $\Omega$ and $\Xi_M$ always come in 
pairs, except possibly for the massless ones. Hence there are no modes with $m^2<0$ for $\Xi_M$. On the other hand, the massless solution 
$\Xi_M^0(z)$ of (\ref{M_mode}) is given by
\begin{equation}
\Xi_M^0(z) \propto e^{3A/2}\frac{\phi'_M}{A'},
\end{equation}
which is however no normalizable since $\Xi_M^0(z)$ is not bounded 
for $z\rightarrow 0$. This is the mode which is expected to be related, 
when gravity is switched off, to the translational zero mode of the flat 
space $SU(5)\times Z_2$ kink 
\cite{Pogosian:2000xv,Vachaspati:2001pw,Pogosian:2001fm}. 
As in the abelian $Z_2$ kinks 
\cite{Kakushadze:2000zn,Shaposhnikov:2005hc}, once we 
include gravity, the massless mode is no normalizable 
and therefore no localized.

The behavior of the scalar fluctuation 
$\Xi_M$ close parallels the one of the scalar fluctuation associated 
to the standard abelian $Z_2$ kink \cite{Giovannini:2001fh}. Now, let 
us consider perturbations along the other generators 
$\mathbf{T}^q$ of $SU(5)$ different from $\mathbf{M}$. For $q\neq M$, 
$\Xi_q(z)$satisfy
\begin{equation}\label{Xi_non_M}
(-\partial_{z}^2 + V_{q})\Xi_q= m^2\Xi_q,
\end{equation}
where $V_q$ is given by
\begin{equation}\label{Vq}
V_{q}= V_{Q_1} + e^{2A}\frac{\partial^2V}{\partial\phi_q^2},
\end{equation}
with $V_{Q_1}$ given by (\ref{VQ1}).

For $q=P$, we find
\begin{equation}
\frac{\partial^2V}{\partial\phi_P^2} = 4b^2(1+ \frac{4}{9}v^2),
\end{equation}
for both symmetry breakings. $V_P$ is everywhere positive 
and hence does not support bound states with $m^2\leq0$. As expected, 
$\Phi$ is stable to perturbations along the $\Phi$ direction. Furthermore, 
these perturbations do not generate normalizable 4D massless modes 
independently of the symmetry breaking pattern.

For $\mathbf{T}^q$ a generator of $H_0$ and $\mathbf{T}^q\neq\mathbf{M}, \mathbf{P}$, we find for the symmetry breaking {\bf A} 
\begin{equation}\label{V_H_0_A}
\frac{\partial^2V}{\partial\phi_q^2}= b^2\left(-8 +\frac{32}{9}v^2 + (3\pm F)(6-\frac{4}{9}v^2\left(F^2 +1\right)\right)
\end{equation}
where $F=F(z)$ is the function defined by
\begin{equation}
F(z)=\tanh b\xi,
\end{equation}
with $\xi=\xi(z)$ (\ref{xi_zeta}). These are perturbations along the six 
generators of $SU(2)_+\times SU(2)_-\subset H_0^{\mathbf A}$, the 
$\pm$ signs correspond to the two different $SU(2)$. On 
the other hand, for the symmetry breaking {\bf B} we find 
\begin{equation}\label{V_H_0_B}
\frac{\partial^2V}{\partial\phi_q^2}= b^2\left(\frac{5}{2} +\frac{109}{36}v^2 + \left(\frac{3}{2} + \frac{1}{6}v^2(5 -\frac{11}{6} F^2)\right)F^2\right)
\end{equation}
for the perturbations along the eight generators of 
$SU(3)\subset H_0^{\mathbf{B}}$. In both symmetry breaking patterns, 
$V_{q}$ is everywhere positive 
and hence does not support bound states with $m^2\leq 0$. From the above 
results, it follows that perturbations along the generators of $H_0$ do 
not generate normalizable 4D massless modes.

\begin{figure}
\centerline{
    \includegraphics[width=0.24\textwidth,angle=0]{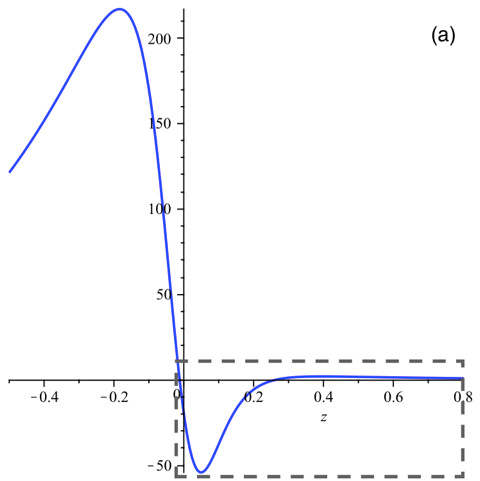}
    \includegraphics[width=0.24\textwidth,angle=0]{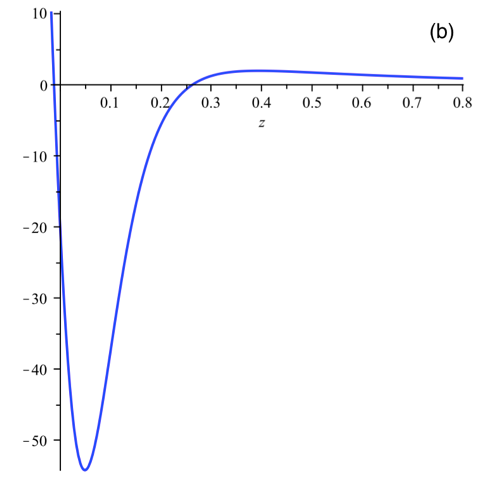}}
\caption{(a) $V_q$ for the $\Xi_q(z)$ modes of the fluctuations within the class described by $H_{-}/H_0$ (for the class described by $H_+/H_0$, $V_q$ is the mirror image). (b) Detail of the area shown in (a).}\label{shape}
\end{figure}

Next, let us consider perturbations along those generators of $SU(5)$ that 
are not generators of $H_0$, which will be called broken generators. For 
a perturbation along $\mathbf{T}^q$ such that $\mathbf{T}^q$ is 
a generator of $H_+$ but not of $H_-$ (plus signs) or a generator of $H_-$ 
but not of $H_+$ (minus signs), we find for the symmetry breaking {\bf A}
\begin{equation}
\frac{\partial^2V}{\partial\phi_q^2}= 2b^2F\left(1 +\frac{2}{3}v^2(1- \frac{1}{3} F^2)\right)(F\pm 1).
\end{equation}
and for the symmetry breaking {\bf B}
\begin{eqnarray}
\frac{\partial^2V}{\partial\phi_q^2}&= &2b^2 F\left( \pm 13(\frac{1}{2}+\frac{4}{9}v^2)\right.\nonumber\\ &&\left. + \left(1 +\frac{2}{3}v^2(1- \frac{1}{3} F^2)\right)(F\pm 1)\right).
\end{eqnarray}
These ones are perturbations along the $n_{\pm}$ generators that 
are broken only at one side of the wall, with $n_{\pm}=4$ for the 
symmetry breaking {\bf A} and $n_{\pm}=6$ for the symmetry 
breaking {\bf B} ($n_{\pm}$ is the dimension of the coset $H_{\pm}/H_0$). 
The Schr\"odinger-like equation for these fluctuations can not be written 
in terms of a fake superpotential as we did for $\Xi_M$. However, on 
general grounds and from the shape of $V_q$ for these perturbations 
(see Fig.\ref{shape}), besides a mild resonance behavior, 
trapped massless modes are expected.

These gravitationally trapped massless modes correspond to rotations 
of $\Phi$ within the class described by $H_{\pm}/H_0$. In the non-gravitating 
setting the analogous zero modes  behave as $F\pm 1$ and are not 
normalizable \cite{Vachaspati:2001pw}. Hence, gravitationally trapped 
massless modes associated to these perturbations can be seen as 
the zero modes in flat space which turn out normalizable and get localized 
once we include gravity. For this type of perturbations there are 
no bound states in the flat space case \cite{Vachaspati:2001pw}, we find 
no modes with $m^2<0$ neither in the gravitating case.   

Finally, for $q$ such that $\mathbf{T}^q$ is not a generator of 
$H_+$ nor of $H_-$, we find that $\partial^2V/\partial\phi_q^2=0$ and 
the modes $\Xi_q(z)$ satisfy
\begin{equation}\label{Xi_q}
(-\partial_{z}^2 + V_{Q_1})\Xi_q= m^2\Xi_q.
\end{equation}
As stated above, $V_{Q_1}$ supports a massless bound state, a tower 
of non-normalizable massive states with $m^2>0$ and no modes 
with $m^2<0$. These ones are perturbations along the $n_{br}$ generators 
that are broken everywhere, with $n_{br}=8$ for the symmetry breaking 
{\bf A} and $n_{br}=2$ for the symmetry breaking {\bf B}.

The above results show that the 5D self-gravitating $SU(5)\times Z_2$ 
domain walls of \cite{Melfo:2011ev} are perturbatively stable.  Additionally, 
on both symmetry breaking patterns, normalizable 4D 
massless modes of the kink fluctuations 
$\boldsymbol{\chi}=\chi_q \mathbf{T}^q$ appear when $\mathbf{T}^q$ is not a 
generator of $H_0$. The existence of normalizable massless 
scalar modes is reminiscent of the situation envisaged in \cite{George:2011tn}, 
although in the latter these modes appears in abelian 
$Z_2$-symmetric models with $N$ scalar fields ($N\geq 2$) and a potential generated by a fake superpotential.
   
\section{Summary and Outlook}

Dynamical localization of gauge fields {\it via} the DS-mechanism 
\cite{Dvali:1996xe} in non-abelian domain wall 
scenarios  has been put forward in 
\cite{Davidson:2002eu,Vachaspati:2003zp} and 
discussed in \cite{Davidson:2007cf} for a flat space model based 
on a $E_6$ symmetry group and in \cite{Melfo:2011ev} for a 
gravitating $SU(5)\times Z_2$ model. Besides the fact 
that the stability of non-abelian kinks is not guaranteed and should 
be proved at least perturbatively \cite{Vachaspati:2001pw,Pogosian:2001fm}, 
it is clear that the 4D phenomenology on the brane may be influenced 
by the inclusion of gravity which couples metric fluctuations with the wall 
fluctuations.

We have thus carried out a diffeomorphism-invariant analysis 
of the fluctuations of the 5D self-gravitating $SU(5)\times Z_2$ domain 
walls of \cite{Melfo:2011ev}, in order to determine their perturbative 
stability as well as their localization properties from the point of view of 
4D observers. 

We find no modes with $m^2<0$ for the fluctuations of the self-gravitating 
$SU(5)\times Z_2$ kinks, implying that these are perturbatively stable as in the flat space-time case \cite{Pogosian:2000xv,Vachaspati:2001pw,Pogosian:2001fm}. 

Not surprisingly, the tensor and vector sectors of the fluctuations behave in 
the same way as the corresponding ones of the standard $Z_2$ kinks. There is 
a normalizable massless mode in the tensor sector which gives rise to 4D 
gravity on the brane and a tower of non-normalizable massive states which 
propagate in the bulk. There is no localized vector fluctuation.
 
On the other hand, the scalar sector of the fluctuations for the 
self-gravitating $SU(5)\times Z_2$ kinks greatly differs from its 
analogous for the abelian $Z_2$ case. We find no normalizable 
4D massless modes associated to the unbroken subgroup $H_0$ on the 
core of the wall, independently of the symmetry breaking pattern 
considered. However, there are as many normalizable 4D zero modes 
as there are broken generators,  i.e., we find gravitationally localized 
massless 4D scalar particles without charge under $H_0$ which can be 
identified as (the 4D zero modes of) the Nambu-Goldstone fields associated 
to the symmetry breaking $SU(5)\times Z_2\rightarrow H_0$. This 
gravitational trapping of Nambu-Goldstone bosons is presumably shared 
with other non-abelian domain-walls, a subject that deserves a further 
separate investigation.

\section{Acknowledgements}
We wish to acknowledge fruitful discussions with Adel Khoudeir, Alejandra Melfo and Aureliano Skirzewski.


\begin{thebibliography}{999}

\bibitem{Randall:1999vf}
  L.~Randall, R.~Sundrum,
  Phys.\ Rev.\ Lett.\  {\bf 83 } (1999)  4690-4693.
  [hep-th/9906064].
  
\bibitem{Guerrero:2002ki}
  R.~Guerrero, A.~Melfo and N.~Pantoja,
  Phys.\ Rev.\ D {\bf 65} (2002) 125010
  [gr-qc/0202011].
  
\bibitem{DeWolfe:1999cp}
  O.~DeWolfe, D.~Z.~Freedman, S.~S.~Gubser, A.~Karch,
  Phys.\ Rev.\  {\bf D62 } (2000)  046008.
  [hep-th/9909134].

\bibitem{Gremm:1999pj}
  M.~Gremm,
  Phys.\ Lett.\  {\bf B478 } (2000)  434-438.
  [hep-th/9912060].
  
\bibitem{Kakushadze:2000zn}
  Z.~Kakushadze, P.~Langfelder,
  Mod.\ Phys.\ Lett.\  {\bf A15 } (2000)  2265-2280.
  [hep-th/0011245].
  
\bibitem{CastilloFelisola:2004eg}
  O.~Castillo-Felisola, A.~Melfo, N.~Pantoja, A.~Ramirez,
  Phys.\ Rev.\  {\bf D70 } (2004)  104029.
  [hep-th/0404083].

  
\bibitem{Bajc:1999mh}
  B.~Bajc, G.~Gabadadze,
  Phys.\ Lett.\  {\bf B474 } (2000)  282-291.
  [hep-th/9912232].
  
\bibitem{Kehagias:2000au}
  A.~Kehagias, K.~Tamvakis,
  Phys.\ Lett.\  {\bf B504 } (2001)  38-46.
  [hep-th/0010112].
  
\bibitem{Ringeval:2001cq}
  C.~Ringeval, P.~Peter, J.~-P.~Uzan,
  Phys.\ Rev.\  {\bf D65 } (2002)  044016.
  [hep-th/0109194].
  
\bibitem{Koley:2004at}
  R.~Koley, S.~Kar,
  Class.\ Quant.\ Grav.\  {\bf 22 } (2005)  753-768.
  [hep-th/0407158].
  
\bibitem{Melfo:2006hh}
  A.~Melfo, N.~Pantoja, J.~D.~Tempo,
  Phys.\ Rev.\  {\bf D73 } (2006)  044033.
  [hep-th/0601161].
  
\bibitem{Liu:2009dw}
  Y.~-X.~Liu, C.~-EFu, L.~Zhao, Y.~-S.~Duan,
  Phys.\ Rev.\  {\bf D80 } (2009)  065020.
  [arXiv:0907.0910 [hep-th]].
    
  
\bibitem{Guerrero:2006gj}
  R.~Guerrero, A.~Melfo, N.~Pantoja, R.~O.~Rodriguez,
  Phys.\ Rev.\  {\bf D74 } (2006)  084025.
  [hep-th/0605160].
  
\bibitem{Huber:2000fh}
  S.~J.~Huber and Q.~Shafi,
  Phys.\ Rev.\  D {\bf 63} (2001) 045010
  [arXiv:hep-ph/0005286].
  
\bibitem{Dvali:2000rx}
  G.~R.~Dvali, G.~Gabadadze and M.~A.~Shifman,
  Phys.\ Lett.\  B {\bf 497}, 271 (2001)
  [arXiv:hep-th/0010071].
  
\bibitem{Dubovsky:2001pe}
  S.~L.~Dubovsky, V.~A.~Rubakov,
  Int.\ J.\ Mod.\ Phys.\  {\bf A16 } (2001)  4331-4350.
  [hep-th/0105243].
  
\bibitem{Batell:2006dp}
  B.~Batell, T.~Gherghetta,
  Phys.\ Rev.\  {\bf D75 } (2007)  025022.
  [hep-th/0611305].
  
\bibitem{Guerrero:2009ac}
  R.~Guerrero, A.~Melfo, N.~Pantoja, R.~O.~Rodriguez,
  Phys.\ Rev.\  {\bf D81 } (2010)  086004.
  [arXiv:0912.0463 [hep-th]].
  
\bibitem{Chumbes:2011zt}
  A.~E.~R.~Chumbes, J.~M.~Hoff da Silva and M.~B.~Hott,
  Phys.\ Rev.\ D {\bf 85} (2012) 085003
  [arXiv:1108.3821 [hep-th]].
  
\bibitem{Germani:2011cv}
  C.~Germani,
  Phys.\ Rev.\ D {\bf 85} (2012) 055025
  [arXiv:1109.3718 [hep-ph]].
  

\bibitem{Dvali:1997sa}
  G.~R.~Dvali, H.~Liu, T.~Vachaspati,
  Phys.\ Rev.\ Lett.\  {\bf 80 } (1998)  2281-2284.
  [hep-ph/9710301].
  
  
\bibitem{Pogosian:2000xv}
  L.~Pogosian, T.~Vachaspati,
  Phys.\ Rev.\  {\bf D62 } (2000)  123506.
  [hep-ph/0007045].
  
\bibitem{Vachaspati:2001pw}
  T.~Vachaspati,
  Phys.\ Rev.\  {\bf D63 } (2001)  105010.
  [hep-th/0102047].
  
\bibitem{Pogosian:2001fm}
  L.~Pogosian, T.~Vachaspati,
  Phys.\ Rev.\  {\bf D64 } (2001)  105023.
  [hep-th/0105128].
  
\bibitem{Vachaspati:2003zp}
  T.~Vachaspati,
  Phys.\ Rev.\ D {\bf 67} (2003) 125002
  [hep-th/0303137].
  
\bibitem{Davidson:2002eu}
  A.~Davidson, B.~F.~Toner, R.~R.~Volkas and K.~C.~Wali,
  Phys.\ Rev.\ D {\bf 65} (2002) 125013
  [hep-th/0202042].

\bibitem{Davidson:2007cf}
  A.~Davidson, D.~P.~George, A.~Kobakhidze, R.~R.~Volkas and K.~C.~Wali,
  Phys.\ Rev.\  D {\bf 77} (2008) 085031
  [arXiv:0710.3432 [hep-ph]].

  
\bibitem{Melfo:2011ev}
  A.~Melfo, R.~Naranjo, N.~Pantoja, A.~Skirzewski and J.~C.~Vasquez,
  Phys.\ Rev.\ D {\bf 84} (2011) 025015
  [arXiv:1104.4857 [hep-th]].

  
\bibitem{Dvali:1996xe}
  G.~R.~Dvali, M.~A.~Shifman,
  Phys.\ Lett.\  {\bf B396 } (1997)  64-69.
  [hep-th/9612128].
  
       
\bibitem{Li:1973mq}
  L.~F.~Li,
  Phys.\ Rev.\  D {\bf 9} (1974) 1723.
  
  
\bibitem{Chen:2010er}
  N.~Chen, T.~A.~Ryttov and R.~Shrock,
  Phys.\ Rev.\  D {\bf 82} (2010) 116006
  [arXiv:1010.3736 [hep-ph]].
  
\bibitem{Giovannini:2001fh}
  M.~Giovannini,
  Phys.\ Rev.\ D {\bf 64} (2001) 064023
  [hep-th/0106041].
  
\bibitem{Callin:2004py}
  P.~Callin and F.~Ravndal,
  Phys.\ Rev.\ D {\bf 70} (2004) 104009
  [arXiv:hep-ph/0403302].
  
\bibitem{Shaposhnikov:2005hc}
  M.~Shaposhnikov, P.~Tinyakov and K.~Zuleta,
  JHEP {\bf 0509} (2005) 062
  [hep-th/0508102].

  
\bibitem{George:2011tn}
  D.~P.~George,
  Phys.\ Rev.\ D {\bf 83} (2011) 104025
  [arXiv:1102.0564 [hep-th]].
  
 
 

  
  
\end{thebibliography}
\end{document}